\documentstyle{article}
\newtheorem{theorem}{Theorem}
\def\a{A}
\def\b{B}
\def\c{C}
\def\d{D}
\def\e{E}
\def\T{T}
\newcommand{\ket}[1]{\left|#1\right>}
\def\ep{\epsilon}
\newcommand{\comment}[1]{}
\begin{document}

\title{
Scalable 
NMR Quantum Computation 
}

\author{Leonard J. Schulman \\
College of Computing \\ Georgia Institute of Technology \\ Atlanta GA
30332 
\and 
Umesh Vazirani\thanks{Supported in part by a JSEP grant.} \\
Computer Science Division \\ U.\ C.\ Berkeley \\ Berkeley CA 94720 }
\maketitle
\begin{center} {\large Preliminary Draft\footnote{Comments welcome to
schulman@cc.gatech.edu and vazirani@cs.berkeley.edu.}
} \end{center}

\begin{abstract}
Nuclear magnetic resonance offers an appealing prospect for
implementation of quantum computers, because of the long coherence
times associated with nuclear spins, and extensive laboratory
experience in manipulating the spins with radio frequency pulses.
Existing proposals, however, suffer from a signal-to-noise ratio that
decays exponentially in the number of qubits in the quantum
computer. This places a severe limit on the size of the computations
that can be performed by such a computer; estimates of that limit are
well within the range in which a conventional computer taking
exponentially more steps would still be practical.

We give an NMR implementation in which the signal-to-noise ratio
depends only on features of NMR technology, not the size of the
computer. This provides a means for NMR computation techniques to
scale to sizes at which the exponential speedup enables
quantum computation to solve problems beyond the capabilities of
classical computers.
\end{abstract}

A sequence of results over the last decade~\cite{DJ, BV, Si, Sh}
have provided the first credible challenge to the widely accepted
notion that all physically ``reasonable'' computer
models are roughly computationally equivalent, i.e.\ a problem is
tractable (e.g.\ solvable in polynomial time) in one model if and only
if it is solvable in another. At issue is the ability of
computers based on quantum physics to perform certain computations
(such as factorization~\cite{Sh}) exponentially faster than
clasical computers.
However, realizing quantum computation 
in the laboratory has proved to be a formidable challenge
since it requires an isolation of the computer from the 
effects of environmentally induced decoherence, while being
able to operate upon its state to perform elementary operations.
Nevertheless, several proposals to realize
quantum computation in the laboratory have been made, using a variety
of systems such as cavity quantum electrodynamics~\cite{DRBH, CY, Tu},
trapped ions~\cite{CZ}, and most recently nuclear magnetic 
resonance~\cite{GC, CFH}. 

The last proposal is particularly interesting
for a number of reasons. Nuclear spins exhibit long relaxation
times --- with coherence times as long as thousands of seconds~\cite{CORW}. 
Moreover, NMR laboratory techniques routinely
manipulate nuclear spins with sequences of hundreds of radio
frequency pulses, and therefore provide a very attractive setting
for carrying out a sequence of computational steps. 
However, for NMR techniques to be useful in
quantum computation, there is a major obstacle that has to
be overcome~\cite{Ll, Di} --- initializing the system in (or near) a
known initial state (say $\ket{0^n}$). By contrast, conventional
NMR systems use macroscopic samples, which at room temperature and in thermal
equilibrium must be regarded as constituting
a statistical mixture of pure states. Of course, if single 
nuclear spins could be individually addressed, this state preparation
problem could be solved~\cite{Wa}. However, this appears to be quite
difficult to realize. 

A major breakthrough in the use of NMR techniques
in quantum computation came about in~\cite{GC, CFH}, where schemes for 
performing small scale NMR quantum computation using bulk samples were first
introduced. The main idea in~\cite{GC}, is to embed a small dimensional
`virtual' pure state within the density matrix describing the bulk
sample, by exploiting the structure present in thermal equilibrium.
Solving the initial state preparation problem in this way paves the 
way for experimental realization of quantum computation using 
off-the-shelf equipment for conventional pulsed NMR. Indeed, this 
approach has been used in the laboratory to implement $2$-qubit
(quantum bit) prototypes of a quantum computer: over $100$ consecutive
logic steps were performed on a $2$-qubit computer, and the basic
steps of Grover's search algorithm \cite{Gr,CGK,CGKL}. Although this
approach provides a 
very important ``proof of concept'' demonstration for quantum
computation, it does not scale --- the strength of signal output by
the NMR quantum computer degrades exponentially in the number of
quantum bits $n$ in the system. Thus the exponential speedup promised
by quantum computation is offset by an exponential increase in the
effort required to detect the output signal. 
The most optimistic predictions are that the output signal will be
undetectable for computers on about $30$ qubits. Quantum computations
of this size could be quite efficiently simulated on conventional
computers.

In this paper, we give a new technique for preparing the initial 
state of the NMR system, where the output signal strength does not 
degrade as the number of qubits in the system is increased. 
We believe this is therefore the first proposal for a quantum computer
which has long decoherence time, scales to large numbers of qubits,
and does not suffer a corresponding decay in signal
strength.

NMR technology requires a ``bulk'' sample in order to create a
readable signal in the output coils. A quantum computer will need to
use enough macromolecules in order to create this signal; this is a
matter for experimental considerations, which we will not discuss here. 
However, we avoid increases in sample size related to the
complexity of the computation; this has the desirable aspect that a
relatively small sample, which may be subjected to extreme 
conditions (of cold, magnetic field, etc.), may offer further
opportunities for increasing the efficiency of the process.

\subsection*{Polarization Process}
Before describing our proposal, we briefly discuss the physical
setting. The outline of the implementation is founded in standard
liquid NMR physics, although as the calculations will indicate,
development of a useful NMR quantum computer will require more.

We consider a collection of macromolecules, each containing $n$ atoms 
with nuclear spin $1/2$ and nuclear magnetic moment $\mu$, suspended in a
liquid medium at temperature $\T$, so that the relaxation (coherence)
time between the particles and the surrounding liquid is on the order
of seconds or thousands of seconds. 

The liquid is subjected to a magnetic field $B_0$. Upon reaching
thermal equilibrium, the difference between the fraction of particles
oriented in the direction of the field, and those oriented in the
opposite direction, is \[ \ep = {\mu B_0 \over kT} \]
where $k$ is Boltzman's constant, approximately $10^{-16}$ in CGS
units. 

A typical magnetic field $B_0$ is approximately
$10^5$ Gauss.
A nuclear magnetic moment such as that of the proton is approximately
$10^{-23}$ in CGS units. 
At room temperature ($\T = 300$~K), with an especially strong
magnet, we can therefore obtain $\ep \approx 3 \times 10^{-5}$.

As will be explained later, the number of qubits upon which a
quantum computation can be performed, is approximately $\ep^2 n$, where
$n$ is the number of spin $1/2$ particles in the macromolecule we
employ as our quantum computer. For $\ep$ in the range obtained above,
and in order to carry out a quantum computation on a useful number of
qubits (e.g.\ $10^2$), this would require an impractically large
macromolecule of size about $10^{11}$.

Hence it is imperative to create a stronger initial polarization
$\ep$. An obvious parameter to consider is temperature. Reducing the
temperature to $10^{-1}$~K gives $\ep \approx 10^{-1}$, therefore
quantum computations on $10^2$ qubits become possible using a molecule
of size about $10^4$. However, it is difficult to obtain
long coherence times at these low temperatures. 

Perhaps a more promising avenue is the use of optical pumping
techniques for boosting the value of $\ep$. Until recently this
technique has been confined to atomic gases, particularly xenon
\cite{BCGHHN,Pines}; values of $\ep$ exceeding $1/2$ have been attained.
There are plans at IBM to explore these techniques for
molecules that may be suitable for quantum computation. With a value
of $\ep$ in this range, the size of the molecules needed for a quantum
computation on $10^2$ qubits would be under $10^3$.

In the remainder of the paper we will simply assume that some $\ep$
has been provided by the polarizing process, and from that starting
point we will show how to initialize the computer so that it can carry
out any desired computation.

\subsection*{Abstract Setting}

We start by describing an abstract computational model that describes
an NMR quantum computer. An ``NMR quantum computer'' is described by
four parameters:  $n,\ell, k$, and $\epsilon$. $n$ is the number of
qubits in the computer (it is the number of spins available for
computation in each molecule of the NMR sample). 

Initially the $n$ qubits are in a thermal mixture which deviates
slightly from a uniform distribution. $\ep$ is the bias induced, at
the start, by the external polarizing process. Namely, if any given
bit of the computer is measured, the probability that $\ket{0}$ is
observed is ${1+\ep \over 2}$.

We assume that the statistical correlation between any two bits on a
molecule, falls off exponentially with the distance between those
bits. $\ell$ is the ``correlation distance'', the distance such that
the correlation falls below some prescribed threshold such as $1/10$. 
We will use the term $\ep$-biased distribution to refer to such a thermal
mixture.

If there were no correlations, the distribution on the bits would be
binomial; in the more realistic case which we consider, we will be
able to obtain all the same essential results as if the distribution
was binomial. Only the analysis will be a little more difficult, and
the numbers a little worse, than for the binomial distribution. 

Why is it sufficient to specify the distribution that results when
we measure the $n$ qubits in the computational basis? To properly
describe the bulk sample in thermal equilibrium, 
we would have to specify
the density matrix associated with the bulk sample. Different
mixtures of pure states with the same density matrix are indistinguishable
by any measurement (so long as that measurement is applied to the
whole ensemble, not to individual members of the ensemble), and
therefore by any quantum computation followed 
by a measurement in the computational basis. However, we will further
restrict the quantum computation that we will allow during the 
state initialization process. The state initialization will be 
carried out by a computation that can only permute the computational
basis states (i.e.\ by essentially a classical computation). 
Under these restrictions, it is sufficient to
specify only the probability distribution
that results when we measure the initial state of the sample in 
the computation basis. This is because different mixtures of pure
states with different density matrices, but with the same resulting
probability distribution, yield the same result under a basis state
permutation followed by a measurement in the basis state. Since at the
end of our initialization process, we plan to obtain $O(n)$ qubits 
in the all $\ket{0}$ state, any further (general) quantum computation
that is restricted to these qubits yields the same results that it
would if started on a $\ket{\bar{0}}$ state.

In addition to the operation of initializing the thermal mixture to an
$\epsilon$-biased distribution, there are four primitive computational
operations that an NMR quantum computer supports:

a) Cyclically shift the $n$ bits clockwise or counterclockwise one
position.

b) Apply an arbitrary two bit operation to the first two bits.

c) Measure the first bit (in some fixed basis).

d) (For a quantum cellular automaton) For some fixed value of $k$
(depending upon the structure of molecule chosen for the NMR
experiment), apply an arbitrary 2-bit operation to all pairs of bits
with indices $lk$ and $lk +1$.

Notes: 1. Operation (a) does not require that the macromolecule have a
cyclic topology. Our operative assumption is a linear topology. The
implementation of the cyclic shift operation is given in the
``Architecture'' section, below. 2. As stated at the outset, these
operations are a model of an NMR quantum computer. 
It must be understood that there is considerable flexibility in the
design of the model, and that for the sake of specificity, we have
made some arbitrary choices; proper choices must eventually be
made on the basis of experimental considerations.

In fact, there can be substantial reward for enriching the above
operations. The machine architecture given by operations (a)-(c)
corresponds to that of a $1$-tape Turing Machine. (We will speak of
the site where we can execute arbitary operations on the pair of bits,
as the ``tape head''.) Later in the paper, after describing designs
which yield operations (a)-(c), we will also briefly describe how a
slight variation of the design can in fact yield the equivalent of a
$2$-tape Turing Machine. (Still on a linear molecule.) With such a
machine, the run time of our algorithm can be significantly improved.

\subsection*{Overall Scheme}

An ideal NMR quantum computer would have its $n$ qubit register
initialized to $\ket{0^n}$. The main goal of this paper is to 
describe an efficient simulation of an ideal NMR quantum computer
using an NMR quantum computer. Notice that if the bias $\epsilon$, 
in the initial state of the NMR quantum computer, were $0$
then the density matrix of the mixture (of the $n$ qubit computers)
would remain unchanged by any sequence of computational steps.
Therefore an NMR quantum computer with parameter $\epsilon =0$ is
incapable of supporting any computation. Our goal is to use the small
but constant bias $\epsilon >0$ to isolate $m = \Theta(n)$ qubits such that the
reduced density matrix of these $m$ qubits is very close to the
density matrix corresponding to the pure state $\ket{ 0^m}$. 

What we need in order to achieve this goal is quite simple: we wish to
carry out a permutation of the computation basis states $x \in \{0,1\}^n$
such that states with low Hamming weight should be reencoded with a long
prefix of $0$'s. A similar task has been addressed previously by a
quantum computation \cite{CD}. However, in that method, the
necessary permutations are accomplished with the aid of a quantum
computer which already has at its disposal a clean workspace, i.e.\ a
sequence of qubits in a known initial state (of size about
$n^{1/2}$). Obtaining such a clean workspace, in an NMR
computer, is precisely the problem which needs to be addressed in
order to make NMR quantum computing possible in the first place. 

In other words, what complicates the construction of these
permutations, for us, is that 
we cannot assume that we have any clean bits at all (i.e.\ bits whose
distribution is almost entirely supported on $\ket{0}$ or $\ket{1}$)
to store intermediate 
results of our computation, since all the available qubits are
in the thermal state. Consequently, and because of the restricted set of
primitive operations allowed on an NMR quantum computer (necessary
because of the physical limitations), we are initially hampered in
the kinds of logical operations we can implement in our computer. 

What we provide is an ``end-to-end'' procedure: we start 
with only a string of qubits in a thermal mixture, and we end with a
string of qubits that with high probability are all in the $\ket{0}$
state. 

\begin{theorem} \label{thm1}
Assume that the thermal mixture is in an $\epsilon$-biased
distribution. Then there is a constant $c$ such that, using primitives
(a) and (b), we can convert the given mixture to one in which $1-o(1)$ of
the probability is supported on strings which begin with a run of
$c \epsilon^2 n$ $0$'s. 
\end{theorem}

The process which we will describe uses $O(n^2)$ steps. 

We will show how to obtain a value of approximately $20$ for $c$.
A slightly more complicated implementation of our method (esp.\ by
using blocks of size greater than $2$ in phase 2, see below) can
decrease this constant further.

\noindent{\bf Proof}

We begin by permuting the bits; if we wish to minimize our reliance
on any assumptions concerning the dependencies among spins in the
original mixture, then the permutation of $\{1,...,n\}$ is chosen at
random, uniformly, by the experimenter. If (as is more likely, and as
was assumed in the previous section) we can assume only local
correlations then it is enough to ``shuffle'' the bits in any
predetermined manner that guarantees that all bits that start out
close to each other (within distance $n^{1/3}$) end up far apart (at
least distance $n^{1/3}$.)  If we can really assume a binomial
distribution on strings, then this initial permutation is unnecessary.
Under weaker assumptions, the permutation is necessary in order for
the probability bounds of the analysis to be valid.

There are a variety of ways to carry out the permutation; using
operations (a) and (b) it 
can be accomplished without difficulty using (to within a constant factor)
the optimum number of transpositions. Typically, and in the worst
case, this number will be on the order of $n^2$.

We will analyze weak (i.e.\ locally correlated) distributions as follows. 
The initialization algorithm has the property that it partitions the 
$n$ bits into blocks of size $n^{1/3}$, and each processed bit output by
the algorithm depends only on one of the blocks. Now, since the $n$
bits were randomly permuted, with high probability no two bits in
any block started out at distance less than $n^{1/3}$.
This implies (even under very weak assumptions on the manner in which
local correlations decay) that the distribution on each block is very closely 
approximated by the binomial distribution. (Under the assumption that
local correlations decay exponentially in distance, the distribution
in the block will have exponentially small distance to the binomial
distribution, in the $L_1$ norm.)

After the initial permutation, we carry out the preparation of the
initial segment of bits. This process will proceed in 
three phases.
\begin{enumerate}

\item Boosting to constant bias: 

In this phase we extract, from $n$ bits with bias $\ep$, 
$\Theta(\ep^2 n)$ bits which have large constant (i.e.\
independent of $n$) bias. This process is efficient (in terms of how
many bits of output are produced) up to a constant factor.

\item Obtaining polynomially small $\delta=(1-\ep)/2$ by increasing
block sizes.

\item Boosting to obtain a nearly perfect block of bits:

In the final phase, while keeping the block size beneath $n^{1/2}$, we
reduce $\delta$ beneath $n^{-10}$. The union bound then implies that
a computation can then begin, working on the assumption that all bits
are $0$'s, and incur only a polynomially small ($n^{-9}$) probability
of error due to possible bad initialization.

\end{enumerate}

\subsection*{Phase 1: Amplification to constant bias}

In phases 1-3 we partition the $n$ bits 
into blocks of size $n^{1/3}$. All computations of phases 1-3 are
conducted internally within these blocks, until after phase 3 the
clean bits are finally collected together in one location for use in a
subsequent computation. In this way we ensure that we can use
near-independence of the bits within each block. If the original
probability distribution was binomial (rather than having local
correlations), there is no need for this device.

{\it {\bf Theorem \ref{thm1}, phase 1: }
Starting with $n$ $\ep$-biased bits, 
and using operations (a),(b), we can with 
probability $1-o(1)$ obtain $\Omega(\ep^2 n)$ bits with bias at least
$0.856$.
}

We will go through several rounds of amplification; as soon as $\ep$
exceeding $0.856$ is achieved, we stop using this process and switch
to phase 2. 

The amplification scheme is very simple. Partition the bits into pairs.
If the bits in a pair are different discard both. Else discard one. 
The expected bias towards $0$ among the surviving bits is $2 \epsilon
\over 1 + \epsilon^2$. 
Also, the expected number of bits that survive is $n {1 + \epsilon^2
\over 4}$. Since the bits are nearly independent (they would be
completely independent if the original distribution was binomial), a
large deviation bound now implies that with probability 
at least $1 - e^{-n/3}$, the number of bits surviving is at least 
${1 \over 4} n - n^{2/3}$.

As we go through several ($k$) rounds, the probability that we wind up
with less than $n 4^{-k} (1-n^{-1/3})^k$ bits is at most $ke^{-n/3}$.
This is negligible.
A little more complicated question is, can we wind up 
with bits with a constant ($0.856$) bias while bounding $4^{-k}$ from
below by $\Omega(\epsilon^{2})$? A positive answer comes from the
following analysis.

 From the
formula $\epsilon_{i+1} = { 2 \epsilon_i \over 1 + \epsilon_i^2}$ we
obtain two things. First, \[\epsilon_i = \epsilon_0 2^i / \prod_{j=0}^{i-1}
(1+\epsilon_j^2).\] So we can rephrase our goal: we wish to upper bound 
$\prod_{j=0}^{i-1} (1+\epsilon_j^2)$ (where
$\epsilon_i=\hat{\epsilon}$). In an ideal process in which $\epsilon$
doubled in each round, we would need $k=\lg
(\hat{\epsilon}/\epsilon_0)$; in the true process we need to increase
$k$ over this ideal quantity by $\lg \prod_{j=0}^{i-1}
(1+\epsilon_j^2)$. In other words, the multiplicative effect on $4^k$
(over the optimal factor), is at most $(\prod_{j=0}^{i-1}
(1+\epsilon_j^2))^2$. 

Second,
\[ \epsilon_i = {1 - \sqrt{1 - \epsilon_{i+1}^2} \over
\epsilon_{i+1}}.\] 
The remainder of this analysis is broken into two parts: the rounds
until $\ep>1/100$, and the remaining rounds until
$\ep>0.856$. 

For the first part we use the inequality
\[ x \leq 0.02 \mbox{ implies }
\sqrt{1-x} \geq 1- {1 \over 2} x - {1 \over 4} x^2 \]
to show that 
\[ \epsilon_i \leq {1 \over 2} \epsilon_{i+1} (1 + {1 \over 2} \epsilon_{i+1}^2).\]
In particular note that this implies \[ \epsilon_i \leq 
0.5004 \epsilon_{i+1} \]
so long as $\epsilon_i$ is beneath our threshold for using this
analysis.

Now, $\prod (1+\epsilon_j^2) \leq e^{\sum \epsilon_j^2}$.
Consequently $\prod (1+\epsilon_j^2) \leq e^{{0.02}^2 {1 \over
1 - 0.5004}}$ 
and so the multiplicative effect on $4^k$ in these rounds
(the factor for how many bits we are losing) is bounded by 
$e^{{0.02}^2 {2 \over 1 - 0.5004}} < 1.0017$.

In the remaining sequence of rounds we have $0.01 < \ep_i \leq 0.856$.
We obtain an upper bound on $(\prod_{j=0}^{i-1} (1+\epsilon_j^2))^2$
by explicitly calculating it beginning with the term corresponding to
$0.856$ and working down, until and including the first term that is
less than $0.01$ (which is the seventh iterate, equal to approximately
$0.009985$). This product is less than $6.7$. \\

\noindent {\bf Implementation: }

We have to be somewhat careful to implement the amplification scheme
using the computational primitives described above. 
We can think of the machine given by primitives (a),(b), as a Turing
machine, whose ``tape head'' is at the site at which arbitrary unitary
operations can be implemented on a pair of adjacent bits. 
We will want to speak of the tape head carrying with it a small
``register'' of several bits: this is easily implemented, by
interspersing rotations of the tape with transpositions at the site of
the ``tape head''. We will use a two-bit register labelled $y_1,y_2$.

We will perform the amplification in stages.
Start with arbitrary
bits in the two-bit register. 
For $m$ ranging from $1$ up to $N/2$
(where $N$ is the current number of bits left in the process ---
initially 
$O(n^{1/3})$), carry out the
two-bit operation ``are they equal?'', namely $\ket{01} \rightarrow
\ket{11}$, $\ket{11} \rightarrow \ket{01}$, on the pair of bits, which
we will call $x_{m,1}, x_{m,2}$.

Now for $m$ ranging from $1$ up to $N/2$, do the following. 
Exchange $x_{m,1}$ with $y_1$, and $x_{m,2}$ with $y_2$. 
Now move the tape head back to the first pair, $x_{1,1},x_{1,2}$. For
$i$ from $1$ to $m-1$, 
do the following: if $y_1=0$, exchange $y_2$ with $x_{i,2}$. Finally,
move the tape head to pair $m$,
and exchange $x_{m,1}$ with $y_1$, and $x_{m,2}$ with $y_2$.

After $m$ reaches $N/2$, and before the next iteration, exchange each
pair of bits $x_{j,1}, x_{2j,2}$ for $1 \leq j \leq N {1+\ep^2 \over
4} (1-o(1))$. This brings all the ``good'' bits to the initial segment
of length $N {1+\ep^2 \over
4} (1-o(1))$. This will be the value of $N$ in the next stage. (The
$1-o(1)$ term, derived from a law of large numbers, 
is chosen so that with high probability all bits in the
segment are in fact ``good'' bits.)

The total number of steps in all stages of all rounds is quadratic in
the block size, hence $O(n^{2/3})$. 

At the end of the process, the $\Theta(n^{1/3} \ep^2)$ good bits lie
in a segment at the start of the block.\\

Why is it necessary to switch to
phase 2 once the bias of the bits is high? Because 
once the bits have high bias, the bit that is discarded in a phase 1
computation itself has substantial bias. Consequently 
the method is wasteful; if we continued with phase 1 to the end, the
ratio of clean bits obtained to the number we started with, would
tend to $0$ in $n$ (rather than being the fixed quantity
$\Omega(\ep^2)$, independent of $n$). In phases 2
and 3 we use blocks that, instead of being of the fixed size $2$,
increase together with the bias. Only one or a constant number of bits
are discarded from each block of the computation, and it becomes
possible to discard a small fraction of the bits, while still
amplifying those that remain.

\subsection*{Phase 2: obtaining polynomially small $\delta$.}

{\it {\bf Theorem \ref{thm1}, phase 2: }
Starting with $n$ bits of bias at least $0.856$, 
and using operations (a),(b), we can obtain $\Omega(n)$ bits 
with $\delta<n^{-0.3}$.
}

This phase will require $O(\log \log n)$ rounds, each using time
$O(n^{2/3})$. 

We begin with $n_0$ bits, of which most are $0$'s, but a constant
fraction, $b_0$, are $1$'s. 

We partition the bits randomly into bins,
each of $k_0$ bits 
$x_1,...,x_{k_0}$. In each bin, we parity bits $x_2$ through $x_{k_0}$ into
bit $x_1$. If $x_1$ equals $1$, we do not pass bits $x_2,...x_{k_0}$
along to the next round; if $x_1$ equals $0$, we do. This is repeated
for several rounds with varying $k$. The bins are
rerandomized in each round. (All the randomness, again, is provided
externally by the experimenter. The computation itself is
deterministic. In particular, all tape movements are oblivious.) \\

\noindent{\bf Analysis: }
Let $\delta_0 = b_0/n_0$. The probability that a given bin contains
exactly one $1$ is (for large $n_0/k_0$) very close to 
\[k_0 \delta_0 (1-\delta_0)^{k_0-1}.\]
(This is what it would be exactly, for independent
sampling with probability $\delta_0$).

Moreover for large $n_0/k_0$, there is a law of large numbers saying
the total number of bins containing one $1$, call it $u$, is with high
probability very close to its expected value, 
\[b_0 (1-\delta_0)^{k_0-1}. \]

{\bf (a)} The total number of bits passed along to the next round,
$n_1$, is lower bounded by only considering bits from blocks which were
entirely $0$'s; this bound (again using a law of large numbers to make
a high-probability statement) is 
\[n_1 \geq {n_0 \over k_0} (1-\delta_0)^{k_0} (k_0-1).\]

{\bf (b)} The total number of $1$'s passed along to the next round is
at most $b_0 - u$ which, w.h.p., is close to its expectation, so
we write 
\[ b_1 \leq b_0(1-(1-\delta_0)^{k_0-1}).\]

Now we need to make a good choice of $k$ as a function of $\delta$.
Note that $\delta=0.072$ corresponds to $\epsilon=0.856$.
Our choice is as follows: for $0.0188 < \delta \leq 0.072$, select
$k=3$. For $0.0027 < \delta \leq 0.0188$, select $k=7$. For $0.000158
< \delta \leq 0.0027$, select $k=21$. For $\delta \leq 0.000158$,
select $k=\delta^{-0.4}$. Note that in this region $k \geq 33$.

In the first of these regions we
are guaranteed $n_1/n_0 \geq 0.532$; in the second we
are guaranteed $n_1/n_0 \geq 0.75$; and in the third we
are guaranteed $n_1/n_0 \geq 0.899$. Each of these regions is
encountered at most once in the process.

In the fourth region, we have $n_1/n_0 \geq {k - 1 \over k}
(1-\delta)^k  \geq e^{-1.1 \delta^{0.4}}$. We also have 
$\delta_1 \leq \delta {1 - (1-\delta)^{k-1} k \over (1-\delta)^k (k-1)}
\leq 1.1 \delta (1 - (1-\delta)^{k-1}) \leq 1.2 \delta^2 (k-1) \leq 1.2
\delta^{1.6}$. 

Consequently, over the entire fourth region, $\prod (n_1/n_0) \geq
e^{-1.1 \sum \delta^{0.4}} \geq 0.96$.

The above iterations halt once we reach a large enough block size,
$n^\alpha$ for 
$0.2 < \alpha \leq 0.32$. At that point we implement another few
iterations using blocks of size $n^{1/3}$ (we can simply use the
entire block of bits that is allowed to interact), bringing $\delta$
down close to the stationary point of the iteration $\delta_1 \leq
\delta_0^2 k$, i.e.\ $\delta=n^{-1/3}$; let us say we halt when
$\delta \leq n^{-0.3}$.

\subsection*{Phase 3: obtaining $\delta < n^{-10}$.}

{\it {\bf Theorem \ref{thm1}, phase 3: }
Starting with $n$ bits of bias at least $1-n^{-0.3}$, 
and using 
operations (a),(b), we can obtain $(1-o(1))n$ bits of bias $1-n^{-10}$.
}

This phase will require a constant number of rounds, each using time
$O(n^{2/3})$ in each $n^{1/3}$-size block, hence $O(n^{4/3})$ time
overall. 

Fix blocks of size $k=n^{1/6}$. Now instead of paritying into just
one bit, parity into the first $2$ bits, i.e.\ compute modulo $4$ the
number of $1$'s in the block.  We now implement the logic gate $(x,y,z)
\rightarrow (x,y,(x \vee y) \oplus z)$ with $x$, $y$ and $z$
representing the first three bits of the block. Now, if after this
gate, the third bit is a $1$, we pass the remaining $n^{1/6}-3$ 
bits of the block on to the next round. Now that the decision has been
encoded in one bit (namely the third bit of the block), this procedure
can be implemented in a manner similar to that described concerning
phase 2 (the ``decision bit'' is carried in the tape head and controls
whether or not a permutation is implemented).

We will only pass $1$'s through to the next round if there are at
least $4$ of them in the entire
block, or any in the first $3$ bits. The recurrence for $\delta$ is
therefore approximately 
\[ \delta_1 \leq \delta_0 (3n^{-1/6} + 3 \delta_0 + {n^{1/6} \choose 3}
\delta_0^{3}) \]
Beginning with the value $\delta \leq n^{-0.3}$ provided by the
previous phase, only a constant number of iterations are required to
reduce $\delta$ beneath $n^{-10}$. The total number of bits is reduced
only by a $1-o(1)$ factor.

\subsection*{Termination}
At this point, in time proportional
to $n^2$, we gather together the remaining bits from all the
$n^{1/3}$-size blocks, ready for a subsequent computation. The probability
that any of these bits are not $0$'s is at most $n^{-9}$.

\subsection*{Efficiency: bit yield}
Collecting together the loss factors from phases 1, 2 and 3, we find
that 
\[{ n_{\mbox{initial}} \over n_{\mbox{final}}} \leq 1.0017 \times
6.7 \times {1 \over 0.532} \times {1 \over 0.75} \times {1 \over 0.899}
\times {1 \over 0.96} \times (1+o(1)) \times \ep^{-2} \leq 20 \ep^{-2} .\]
This factor can be improved by using more complex computations.
The chief place to obtain gains is in the latter stages of
phase 1 and the earlier stages of phase 2; in both cases the way to
improve efficiency is to use larger block sizes, and a more
complicated permutation within each block, in order to extract a
fraction of bits from the block that tends to the optimal fraction,
$({\ep_{i} \over \ep_{i+1}})^2$. 

\subsection*{$\Omega(\epsilon^2 n)$ clean bits is optimal}
It was noted above that if $\epsilon=0$, we cannot prepare any bits
at all that are biased toward $\ket{0}$. If $\epsilon > 0$, how
many such bits can we hope to prepare? If we ask that with high
probability $k$ bits are all $0$'s, then the central limit theorem
places a 
limit on $k$ of $n(1-H_2({1+\epsilon \over 2}))$ which, for small
$\epsilon$, is approximately $n \epsilon^2$. 
To prepare just one good bit, therefore, we must use about
$\epsilon^{-2}$ bits with bias $\epsilon$. 

\section*{Architecture}

We now discuss how the computational primitives (a),(b), and some
extensions, can be implemented on polymers with certain kinds of
periodic structures.

\subsection*{Turing machine: }

Normally one imagines a Turing machine having a ``head'' which
implements computations locally, i.e.\ involving the state of the ``tape''
in the vicinity of the head. We implement this abstraction (but
without any moving parts) in the following way. (It must be understood
that there is considerable flexibility in the design, and 
that for the sake of specificity, we are making some arbitrary choices;
the proper choices must eventually be made on the basis of experimental
considerations.)

The tape will not of course be infinite, but a ring of $n$ qubits.
These will be realized in the nuclear spins of a linear polymer. The
polymer will consist of $n/3$ repetitions of the sequence $\a\b\c$, thus
$\a\b\c\a\b\c\a\b\c\a\b\c...$; the atoms $\a,\b,\c$ have spin $1/2$
nuclei. In addition, at one point in the chain, 
another atom, $\d$, is adjacent to the chain, near a neighboring pair
of $\c$ and $\a$ atoms; it induces a chemical shift in some of the
energy levels at these two neighboring atoms. 

(Note: it is not actually necessary for $\a$,$\b$ and $\c$ to be
different types of nuclei; they could all be of one kind, if the
periodic structure resides in adjacent atoms that induce suitable
chemical shifts in the energy levels.)

Five resonant frequencies will be such that we can
implement the following five operations:

\begin{enumerate}
\item Frequency 1: transposition of the qubits in all adjacent $\a\b$
pairs. 
\item Frequency 2: transposition of the qubits in all adjacent $\b\c$
pairs. 
\item Frequency 3: transposition of the qubits in all adjacent $\c\a$
pairs.
\item Frequencies 4,5: these resonate only with energy levels shifted
by the presence of atom $\d$. Hence they induce a unitary operator
only on the pair of qubits at the $\c$ and $\a$ atoms immediately
adjacent to atom $\d$. We assume that the combinations of frequencies
4 and 5 generate the group of all transformations in that
$4$-dimensional Hilbert space. 
\end{enumerate}

Arbitrary ``oblivious'' quantum computations can be performed on this
machine. By an ``oblivious'' computation we mean one in which the
sequence of movements of the tape head is a function is the same in
all the superposed ``copies'' of the machine, in the quantum
computation.

A cyclic shift of the tape by one position is implemented by the
following sequence of transpositions: $(\a,\b)$, $(\c,\a)$, and then
$(\b, \c)$.
(Each such transposition can be
implemented by three CNOT gates: for example $(\a,\b)$ can be
implemented by the sequence $[\a \rightarrow \b], [\b
\rightarrow \a], [\a \rightarrow \b]$.) 
A succession of such triples of transpositions will bring any desired
pair of adjoining qubits next to the tape head.

\subsection*{Cellular automaton with distinguished site: } 

Lloyd\cite{Ll} has proposed implementing a quantum cellular automaton.

We propose an architecture similar to what we have described above,
but now we use five kinds of atoms: three $(\a,\b,\c)$ have spin $1/2$
nuclei and two $(\d,\e)$ induce chemical shifts in resonant
frequencies of nearby atoms of the first three types. We assume that
$k | n$. The ring consists of repetitions of the pattern $\a \b \c$;
after every $k$ atoms of type $\a,\c$, one atom of type $\d$ adjoins
the chain and induces local chemical shifts. At one site an $\e$ atom
adjoins the chain and induces chemical shifts, which are different
from those induced by $\d$.

One step of the computation is implemented by a pulse at a frequency
that involves a $\d$ atom and the two adjacent spin $1/2$ atoms;
rotations of the tape are implemented as above, small rotations 
allow information to be sent between adjacent ``cells'' of the
cellular automaton, while global rotations bring the tape contents
past the $\e$ site, where individual operations may be implemented.

\subsection*{Two-tape Turing machine: }

To implement a two-tape Turing machine we need to enable the head to
move independently on each of the tapes. Equivalently, in our
implementation, we need to have two cycles of bits, which can
independently be cyclically shifted past the head.

Let the molecule consist of $n$ repetitions of the sequence
$\a\b\c\d$. (As above, these are spin $1/2$ nuclei and each adjacent
type of pair can be addressed with distinctive frequencies.)

The $\a$ and $\c$ nuclei will carry one tape, the $\b$ and $\d$
nuclei the other. (Note that the nuclei of any given type carry a
contiguous segment of half a tape, not every other bit.)

The sequence of transpositions $(\a\b)(\b\c)(\a\b)(\c\d)(\a\d)(\c\d)$
rotates the $\a\c$-tape by one position, while leaving the $\b\d$-tape
fixed.

The most time-consuming stages of our procedure are the initial
permutation of the bits and the final collecting of the clean bits,
each requiring time $O(n^2)$. In fact, these are the only stages which
require more than time $O(n^{4/3})$.

The terminal permutation is very simple; the initial permutation can
be very simple, as well, so long as we make the ``local correlations''
assumption on our initial $\ep$-biased distribution, in which case we
can use the permutation which sends bit $r n^{1/3} + s$ (for $0 \leq s
< n^{1/3}$) to position $(r+s)n^{1/3}+s$.  In this case, the initial
permutation can be performed in time $O(n^{4/3})$, and the final
permutation in linear time, on the $2$-tape
architecture. Consequently, the entire procedure can be implemented in
time $O(n^{4/3})$. If we further augment our device by combining the
features of a $2$-tape machine with those of a cellular automaton,
with $k=n^{1/3}$, then the initial permutation can be performed in
linear time, and in phase 3 and the latter part of phase 2 we
can gain time by working in parallel within each $n^{1/3}$-size
block. The overall runtime reduces to linear.

Thus there is substantial benefit in implementing slightly
stronger primitives than the minimal list of operations (a)-(c).

\subsection*{Acknowledgments}
Thanks to Isaac Chuang and Richard Singerman for helpful discussions.

\end{document}